**Authors**
Yusuke Nakazawa*, Takafumi Akiho, Kiyoshi Kanisawa, Hiroshi Irie, Norio Kumada,
and Koji Muraki

**Affiliation**
NTT Basic Research Laboratories, NTT Corporation, 3-1 Morinosato-Wakamiya,
Atsugi 243-0198, Japan

**E-mail**
yusuke.nakazawa@ntt.com

**Abstract**

Here, we study the effects of a GaAs buffer layer on the structural, magnetic, and transport properties of $Cr_y(Bi_xSb_{1-x})_{2-y}Te_3$ magnetic topological insulator thin films and compare them with those of $V_y(Bi_xSb_{1-x})_{2-y}Te_3$, which we recently reported. Similar to the case of $V_y(Bi_xSb_{1-x})_{2-y}Te_3$, growth on a GaAs buffer layer leads to some distinctly different properties than direct growth on InP substrates. These include improved interface quality confirmed by transmission electron microscopy, enhanced magnetic coercive fields, and smaller resistivity peaks at the magnetization reversals. Furthermore, the Bi-ratio dependence of the carrier density reveals that the interface property also affects the Fermi level. These results demonstrate the importance of the buffer layer in controlling the electronic properties of the magnetic topological insulator films.

**1. Introduction**

The magnetic topological insulator (MTI) is a class of topologically nontrivial states of matter with spontaneously broken time-reversal symmetry[1,2]. Among the unique properties of the MTIs, the quantum anomalous Hall (QAH) effect, characterized by quantized Hall resistance $R_{yx} = \pm h/e^2$ in the absence of external magnetic fields[3,4], is one of the representative phenomena. After the first observation in a magnetically doped topological insulator[5], the QAH effect has been observed in various material systems, including intrinsic magnetic topological insulator[6], bilayer graphene[7], and transition metal dichalcogenide[8]. Among them, $Cr_y(Bi_xSb_{1-x})_{2-y}Te_3$ (CBST) and $V_y(Bi_xSb_{1-x})_{2-y}Te_3$ (VBST) have been the most studied systems[5,9]. Their surface energy gap becomes nontrivial when the exchange energy due to spontaneous magnetization exceeds the hybridization energy of the top and bottom surfaces, where gapless chiral edge states emerge inside the gap. They have been studied both from a physical point of view, such as dissipation dynamics[10–12] and universal scaling of the quantum phase transition[13–18], and from an application point of view, such as the resistance standard[19,20] and spintronics[21,22].

In terms of sample preparation, increasing the observation temperature of the QAH effect has been an important issue, where the observation temperature is typically below ~ 0.1 K[23–25] while their Curie temperature is on the order of 10 K[5,9]. In previous reports, modified layer structures using magnetic modulation doping[23] and a magnetic capping layer[25] have been shown to be effective in raising the observation temperature. However, these samples were grown directly on substrates, such as $SrTiO_3$(111)[5,9,10,17,22] and InP(111)A[14,20,23], and little was known about the effects of a buffer layer on the properties of MTI films.

Recently, we have studied the effects of a GaAs buffer layer on the structural and transport properties of VBST films compared to samples grown directly on InP substrates[26]. We have found that the VBST grown on the GaAs buffer is distinct from that grown on the InP substrate in three aspects. First, the crystal and interface quality is improved by implementing the GaAs buffer. Second, in addition to the structural property, magnetic property is also changed; the coercive field $B_c$ is increased to almost double on the buffer. In particular, $B_c$ = 1.34 T of our sample on the GaAs buffer is the largest among those reported for VBST. Third, the resistivity peak value at the magnetization reversals is smaller on the buffer, suggesting that the dimensionality of the QAH state is affected by the interface quality. These observations in VBST have suggested that the buffer layer is important in engineering the QAH insulator films and their heterostructures with controlled structural and magnetic properties.

Following these results in VBST, here, we study the effects of the GaAs buffer layer on CBST films compared to those grown directly on InP substrates. This is important, for example, because VBST and CBST have significantly different coercive fields, so the combination of them allows otherwise unavailable magnetization configurations in their heterostructures[27,28]. Furthermore, to gain deeper insight into the effects of the GaAs buffer layer, we compare the Bi ratio dependence of the magnetic and transport properties of the CBST films. Scanning transmission electron microscopy (STEM) confirms that the GaAs buffer is also effective in improving the interface quality of the CBST films. The GaAs buffer layer also affects the magnetic and transport properties of CBST films in a manner similar to VBST; enhanced coercive fields and smaller resistivity peaks are observed at the magnetization reversals. Interestingly, conductivity scaling shows that, by using the GaAs buffer, the electronic states in both CBST and VBST are more three dimensional than those on the InP substrate despite the same physical thickness, suggesting enhanced exchange coupling. We also compare the Bi-ratio dependence of the coercive fields, carrier density, and anomalous Hall resistivity of the CBST films on the GaAs buffer and InP substrates. These comparisons suggest that the optimum Bi composition, which gives the highest anomalous Hall resistivity and the lowest carrier density, differs depending on the buffer layer or substrate, confirming that the Fermi level is also affected by the interface property.

## 2. Experiments

We used a multi-chamber MBE system to grow the GaAs buffer and MTI layers in separate growth chambers connected by an ultra-high vacuum transfer line. The samples were grown using the same procedures described in our previous work[26]. We used an on-axis GaAs(111)A substrate, for which a 200-nm-thick GaAs buffer layer with an atomically flat surface can be obtained under optimized growth conditions[29–31]. For CBST, the growth temperature $T_g$ was 170°C, which was measured by a thermocouple. The Cr ratio $P_{Cr}/(P_{Sb} + P_{Bi} + P_{Cr})$ was set to 0.027 (i.e. $y = 0.054$), where $P_{Sb}$, $P_{Bi}$, and $P_{Cr}$ are the beam-equivalent pressures for each element[32]. For VBST, the growth condition was slightly different; $T_g = 150°C$ - 160°C and $P_V/(P_{Sb} + P_{Bi} + P_V) = 0.025$ (i.e. $y = 0.050$). The Bi ratio $x = P_{Bi}/(P_{Sb} + P_{Bi})$ is between 0.35 to 0.70 in this study. We also grew CBST and VBST directly on semi-insulating InP(111)A substrates as control samples[33]. For both samples grown on the GaAs buffer and the InP substrate, the thickness of the CBST and VBST layers is 6 or 7 quintuple layers (QL, 1 QL ≑ 1 nm), which was determined from the period of the Laue oscillations in X-ray diffraction (XRD)[26] and consistent with that estimated from the growth rate (approximately 0.08 QL/min). Capping layer was not used in this study. Reflection high-energy electron diffraction and atomic force microscopy confirmed that the surface of the CBST and VBST layers is two-dimensionally flat. Single-crystalline nature of the samples is confirmed by XRD $\theta$-$2\theta$ scans as shown in **Figure S1** in the supporting information. Detailed structural properties, especially their interface quality, were characterized by STEM and energy-dispersive X-ray spectroscopy (EDX). Transport measurements were performed at 500 mK in a $^3$He refrigerator for the CBST and at 20 mK in a dilution refrigerator for the VBST, with magnetic fields up to 3 T. The Hall and longitudinal resistivities ($\rho_{yx}$ and $\rho_{xx}$) were measured with the samples cut into 1-mm wide pieces and Ohmic contacts made with silver paint.

## 3. Results

First, we present STEM characterization of the CBST samples to investigate their interface quality. **Figure 1**(a) and 1(b) are cross-sectional high-angle annular dark field STEM (HAADF-STEM) images of the CBST samples grown on the GaAs buffer layer and InP substrate, respectively. Notably, the interface between the CBST and GaAs buffer layer is much sharper with no transition region, while there is an inhomogeneous intermediate layer between the CBST and InP substrate. EDX indicates that Cr is accumulated at the inhomogeneous layer as shown in **Figure S2** in the supporting information. The Cr accumulation may be caused due to incomplete oxide removal from the substrates, as we have discussed previously for the V accumulation in VBST on InP[26]. Such unintentional Cr accumulation at the substrate interface has been commonly observed in previous reports[34,35]. Figure 1(c) and 1(d) are depth profiles of the in-plane lattice constants calculated from the Fourier transform of the STEM images. The in-plane lattice constant changes abruptly at the interface with the GaAs buffer, while a gradual change is observed across the Cr-accumulated region on the InP substrate, which also represents their distinct interface quality. These results demonstrate that the implementation of the buffer layer is effective for obtaining an improved interface free from interfacial accumulation for both CBST and VBST.

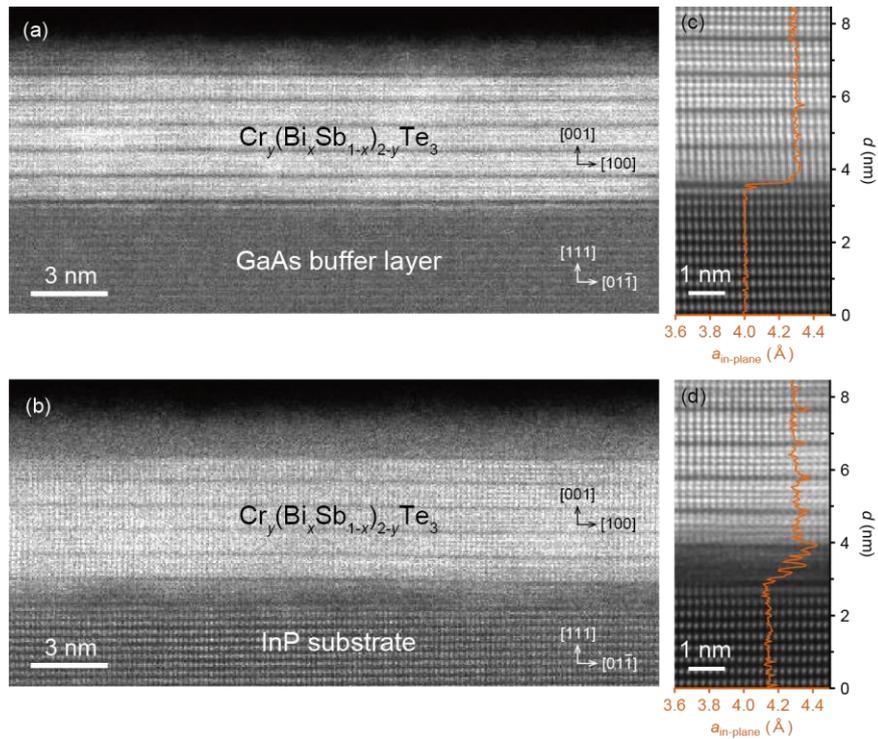

**Figure 1**: (a), (b) Cross-sectional HAADF-STEM images for the CBST films (a) grown on the GaAs buffer layer ($x = 0.55$) and (b) directly on the InP substrate ($x = 0.50$), respectively. (c), (d) Depth profiles of in-plane lattice constants $a_{\text{in-plane}}$ calculated from the Fourier transform of the STEM images.

Next, we present the results of the transport measurements. We measured the Hall and longitudinal resistivity ($\rho_{yx}$ and $\rho_{xx}$) of the CBST and VBST films grown on the GaAs buffer and InP substrates as a function of the magnetic field [**Figure 2**(a) - 2(d)]. The VBST samples exhibit the QAH effect ($\rho_{yx} = \pm h/e^2$ and $\rho_{xx} = 0$) on both the GaAs buffer and the InP, as we have previously reported[26]. Due to the higher measurement temperature used here, the Hall resistivities of the CBST samples are not quantized, which in turn highlights the difference, i.e., the higher Hall resistivity, 80% of the quantization value, is observed on the GaAs buffer compared to 48% on the InP substrate. Notably, we observe the enhancement of the coercive fields $B_c$ in both CBST and VBST. As we have previously discussed for VBST [26], the difference in the coercive fields may be due to the interfacial accumulation of the magnetic dopants on the InP substrate, which may reduce the actual dopant composition in the bulk of the CBST and VBST layers.

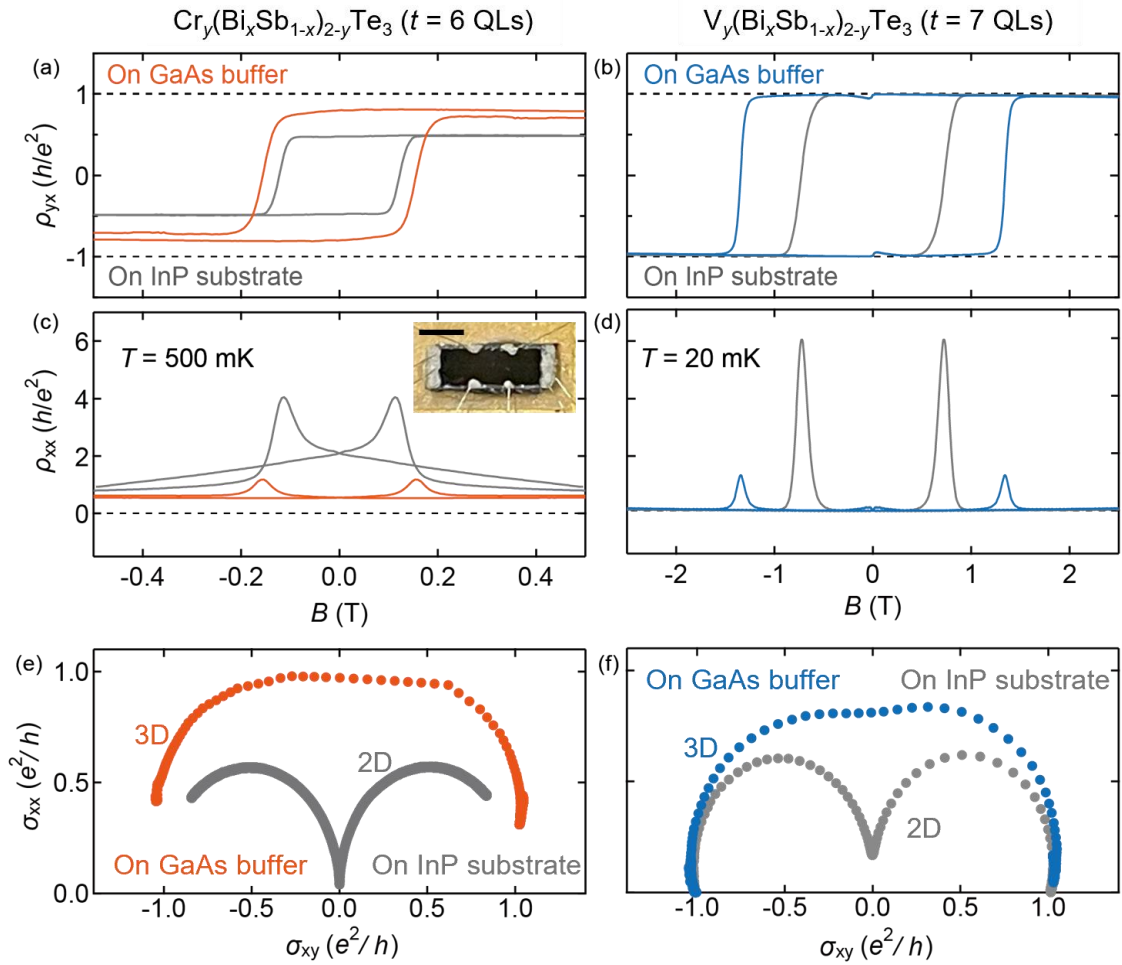

**Figure 2**: (a), (b) Hall ($\rho_{yx}$) and (c), (d) longitudinal ($\rho_{xx}$) resistivity as a function of magnetic fields and (e), (f) scaling diagrams of ($\sigma_{xx}$, $\sigma_{xy}$) of the CBST ($x = 0.55$) and VBST ($x = 0.35$) films grown on the GaAs buffer layer and the InP substrate. The inset in (c) is an image of the sample used for the transport measurement. The black bar corresponds to 1 mm.

In addition to the coercive fields, the dimensionality of the QAH states also differs between the samples grown on the GaAs buffer and the InP substrates. For CBST, the resistivity peaks at the magnetization reversals are smaller on the GaAs buffer [Figure 2(c)], similar to the case for VBST [Figure 2(d)] that we reported previously. Another report on VBST that examined the dimensionality of the QAH insulator has shown that thinner (thicker) samples exhibit higher (lower) resistivity peaks[18]. Given this, our results suggest that the electronic states in both CBST and VBST are more three dimensional than those grown directly on InP substrates, despite the same physical thickness. The difference in the dimensionality is highlighted by analyzing the conductivity scaling. Figure 2(e) and 2(f) show the scaling diagrams of conductivity ($\sigma_{xx}$, $\sigma_{xy}$) for the CBST and VBST, respectively. For both the CBST and VBST, the scaling behavior is clearly different between those grown on the GaAs buffer and the InP substrate; a single semicircle centered at (0, 0), indicating a three-dimensional scaling, is observed on GaAs, while double semicircles centered at (0, $\pm e^2/2h$), indicating a two-dimensional scaling, is observed on InP[18]. The difference in the dimensionality is also suggested in temperature dependence of the transport properties (**Figure S3** in the supporting information). The different dimensionality can also be understood with the aforementioned reduced dopant composition on the InP substrates due to the interfacial accumulation. The dimensionality of the electronic states in an MTI film is determined by the interplay between the hybridization energy of the top and bottom surfaces and the exchange energy, with a smaller exchange energy making them more two dimensional.

To gain further insight into the effects of the GaAs buffer layer, we compare the Bi ratio dependence of the magnetic and transport properties of the CBST films. **Figure 3**(a) plots the coercive fields measured at 1.5 K as a function of the Bi ratio $x$. A nearly twofold enhancement of $B_c$ is consistently observed for the $x$ studied for the CBST grown on the GaAs buffer. We also observe a decrease in $B_c$ with increasing $x$ for $x \leq 0.6$, both for the samples on the GaAs buffer and the InP substrates. Note that the long-range magnetic order in CBST is mediated by the interaction between Sb $p$ and Cr $d$ states[36,37], which explains the lower $B_c$ observed for larger Bi (i.e. smaller Sb) ratio up to $x \leq 0.6$. On the other hand, for the CBST grown directly on the InP substrates, $B_c$ increases with further increasing $x$ above 0.6. This contrasts with the previous report for bulk CBST[37], where $B_c$ decreased significantly for $x = 0.6$ and 0.8 due to the intermixing of different magnetic phases. Thus, our results may represent the behavior of a single magnetic phase obtained by the non-equilibrium growth characteristic of MBE.

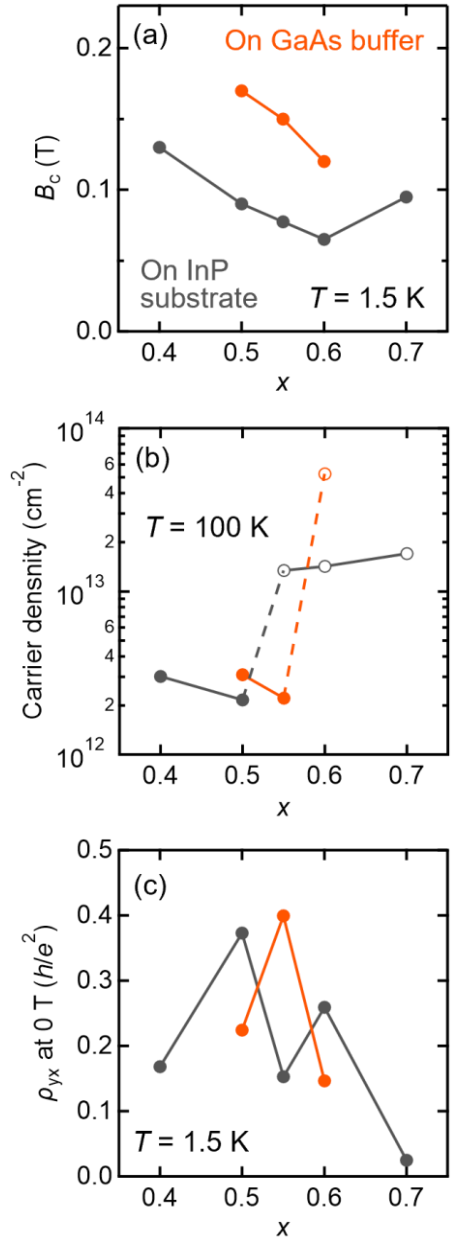

**Figure 3**: (a) Coercive field $B_c$, (b) carrier density, and (c) anomalous Hall resistivity of the CBST films ($t = 7$ QLs) grown on the GaAs buffer layer and the InP substrate. Coercive field and anomalous Hall resistivity are measured at 1.5 K, while the carrier density is measured at 100 K. The closed (open) symbols in (b) represent p-type (n-type) carrier.

Figure 3(b) shows the carrier density of the CBST films. We evaluated the carrier density from the slope of the Hall resistivity at 100 K to minimize contributions from the anomalous Hall component (**Figure S4** in the supporting information). We notice that the Bi ratio for the minimum carrier density shifts to the Bi-rich side for the samples grown on the GaAs buffer layer, suggesting that the Fermi level is affected by the interface property. This shift can also be understood by the scenario of altered Cr composition. In CBST, the Cr ions act as hole donors[37]. Thus, the optimum Bi ratio for charge neutrality can differ between samples on the GaAs buffer and the InP substrates if the actual Cr composition in the bulk of the CBST layers is different, as mentioned above.

Figure 3(c) shows the anomalous Hall resistivity measured at 1.5 K, which is defined as $\rho_{yx}$ at 0 T after saturating the magnetization. Note that a previous report on VBST has shown that the Bi composition dependence of the anomalous Hall resistivity at 4.2 K follows the same trend as that at $T \leqq 50$ mK[38]. This suggests that the optimum Bi composition to observe the QAH effect at low temperature can be determined from the measurements at higher temperatures. Notably, for both samples on the GaAs buffer and the InP substrate, the anomalous Hall resistivity is maximum at the Bi ratio for the minimum carrier density. This also indicates that tuning the Bi-Sb ratio to locate the Fermi level at charge neutrality, in accordance with the interface properties is important to obtain a CBST sample with a large anomalous Hall resistivity.

## 4. Conclusion

In summary, we studied the effects of a GaAs buffer layer on the characteristics of the magnetic topological insulator CBST. STEM EDX analysis demonstrated that the implementation of the buffer is effective in overcoming the interfacial accumulation of magnetic dopant. We also evaluated the effects of the buffer on the magnetic and transport properties of CBST. The suppression of interfacial accumulation enhances the coercive field and exchange energy, the latter being reflected in more three-dimensional transport behavior. In addition, the Bi ratio dependence of the carrier density of CBST suggests that the Fermi level is also changed by the implementation of the buffer layer. Together with our previous study on VBST, these results further confirm that the buffer layer is important for controlling the structural and magnetic properties of MTI films and will be desirable for engineering QAH insulators with improved interface and controlled band structure.

**Supporting Information**

Supporting Information is available from the Wiley Online Library or from the author.

**Conflict of Interest**

The authors have no conflicts to disclose.

Supporting information

# Effects of GaAs Buffer Layer on Structural, Magnetic, and Transport Properties of Magnetic Topological Insulators $Cr_y(Bi_xSb_{1-x})_{2-y}Te_3$ and $V_y(Bi_xSb_{1-x})_{2-y}Te_3$ Films

Yusuke Nakazawa*, Takafumi Akiho, Kiyoshi Kanisawa, Hiroshi Irie, Norio Kumada, and Koji Muraki

NTT Basic Research Laboratories, NTT Corporation, 3-1 Morinosato-Wakamiya,
Atsugi 243-0198, Japan

*yusuke.nakazawa@ntt.com

**S1. X-ray analysis of the CBST films**

Figure S1 shows XRD $\theta$-$2\theta$ scans for the CBST films grown on the GaAs buffer layer and the InP substrate. For both samples, diffraction peaks from the (003) series of CBST are clearly visible, confirming their single-crystalline nature. On the other hand, the Laue oscillations around the diffraction peaks are more pronounced for the sample grown on the GaAs buffer layer, showing its better flatness. The CBST thicknesses derived from the oscillation periods are 6.1 and 5.9 nm for the samples on the GaAs buffer and the InP substrate, respectively, which are consistent with the designed thickness of six quintuple layers.

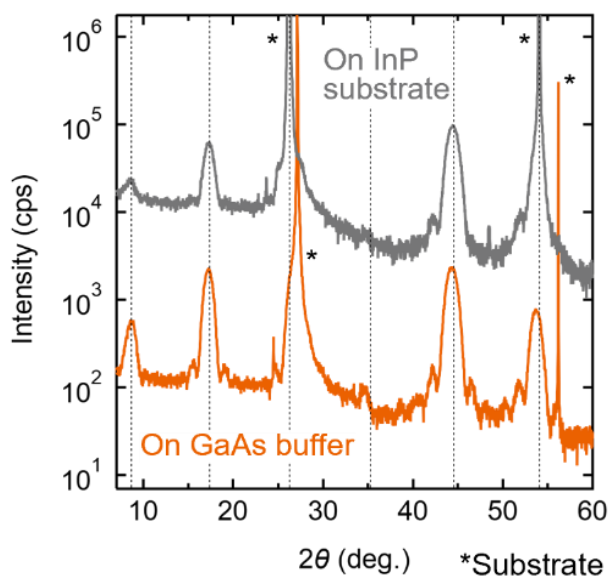

**Figure S1**: XRD $\theta$-$2\theta$ scans of the CBST films grown on the GaAs buffer layer and the InP substrate. The dashed lines show diffraction peak positions for the CBST(003) series. The asterisks denote diffraction peaks from the substrates. The plots are vertically offset for clarity.

## S2. STEM-EDX analysis of the CBST films

Figure S2 shows the STEM-EDX line profiles of Cr for the CBST films grown on the GaAs buffer layer [Figure S2(a)] and the InP substrate [Figure S2(b)]. For the sample grown on the InP substrate, the Cr profile shows a peak at the interface between the CBST and the substrate, suggesting the interface accumulation of Cr. In contrast, the Cr profile for the CBST grown on the GaAs buffer layer does not show such a peak, indicating the absence of interface accumulation.

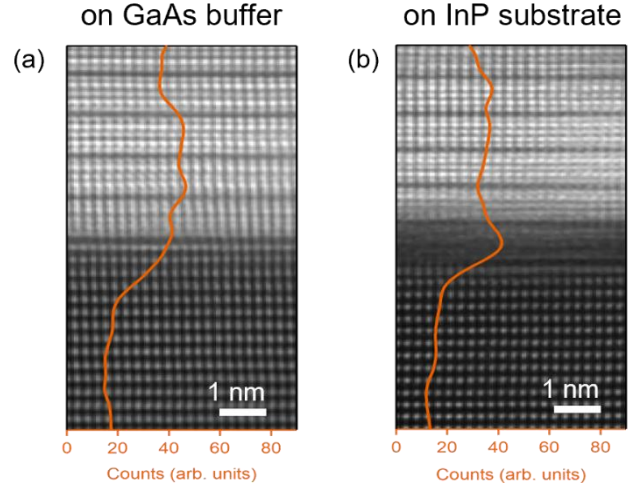

**Figure S2**: STEM-EDX line profiles of Cr for the CBST grown on (a) the GaAs buffer layer and (b) the InP substrate.

## S3. Temperature dependence of transport properties

Figure S3(a) – (h) show longitudinal resistivity $\rho_{xx}$ and scaling diagrams of the CBST and VBST samples shown in Figure 2 in the main text, measured at different temperatures. Figure S3(i) and S3(j) show the temperature dependence of the $\rho_{xx}$ peak values for the CBST and VBST samples, respectively. For both CBST and VBST grown on the GaAs buffer, the $\rho_{xx}$ peak height is nearly temperature independent with a value close to $h/e^2$. On the other hand, the $\rho_{xx}$ peak value is higher than $h/e^2$ and increases with lowering temperature for the samples grown directly on the InP substrate. These behaviors can be compared to the previously reported thickness-dependent study for VBST, where the $\rho_{xx}$ peak value is nearly temperature independent in the three-dimensional limit (thick films), while it shows a thermally activated high-resistance behavior in the two-dimensional limit (thin films)[S1]. The different temperature dependence we observed for the $\rho_{xx}$ peak value supports that the dimensionality of the electronic states in MTI films differs between samples grown on the GaAs buffer and the InP substrate.

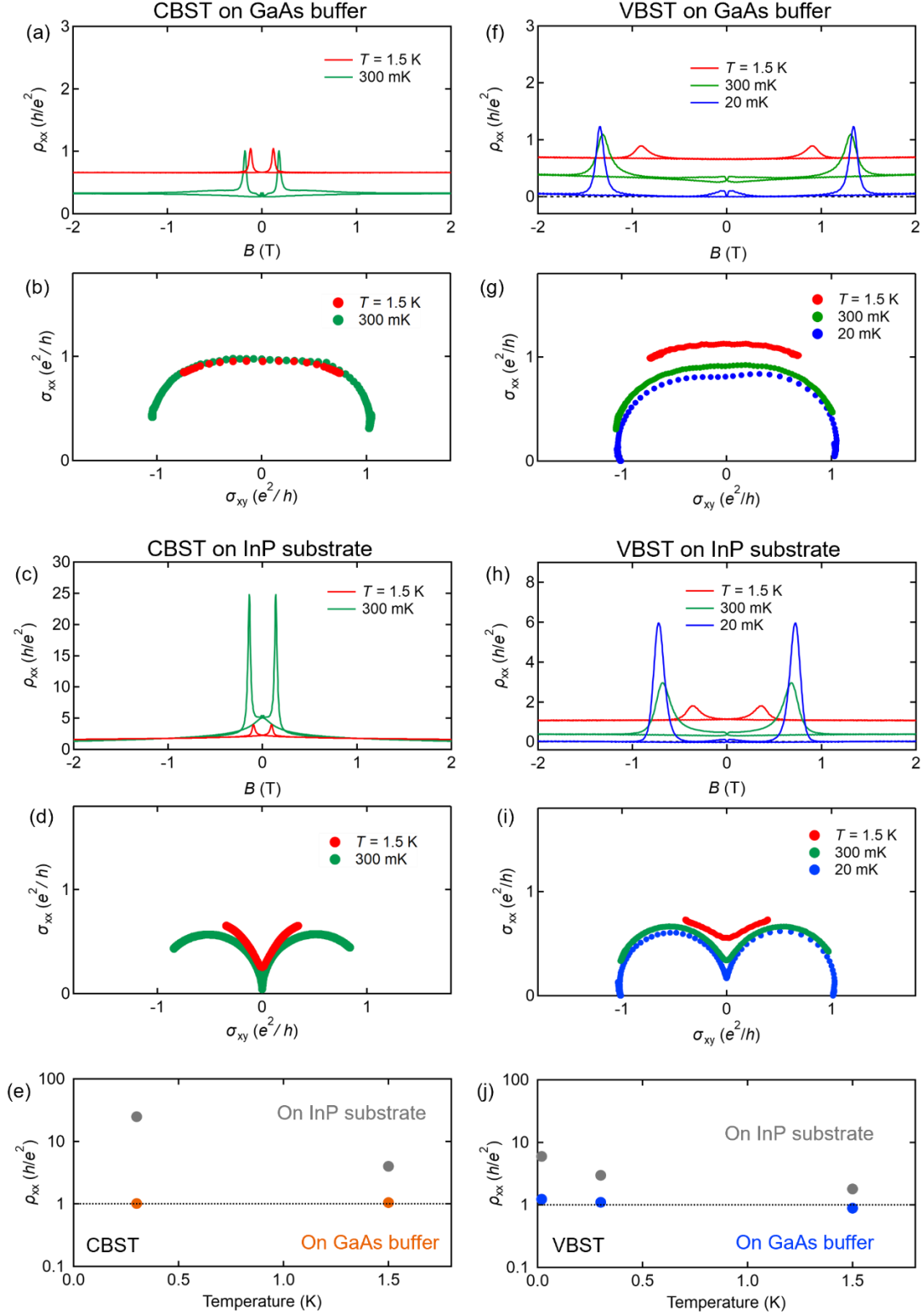

**Figure S3:** Magnetic field dependence of the longitudinal resistivity $\rho_{xx}$ and the scaling diagrams for (a) and (b) CBST on the GaAs buffer, (c) and (d) CBST on the InP substrate, (f) and (g) VBST on the GaAs buffer, and (h) and (i) VBST on the InP substrate, measured at different temperatures. (e) and (j) $\rho_{xx}$ peak values for (e) CBST and (j) VBST plotted as a function of temperature.

## S4. Hall measurements at 100 K

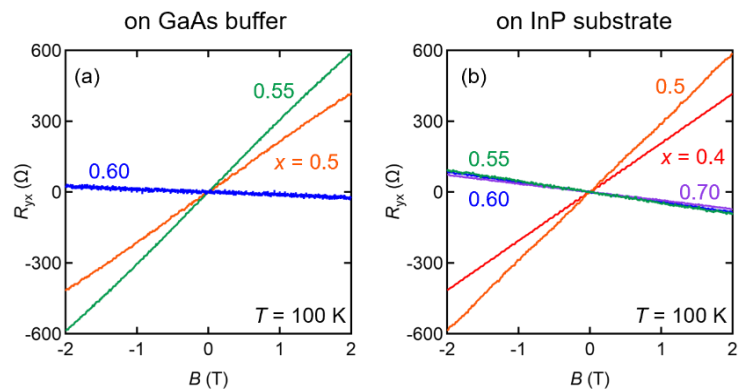

**Figure S4**: Hall measurement data of the CBST films grown on (a) the GaAs buffer layer and (b) the InP substrate, which are used to calculate the carrier density shown in Figure 3(b) in the main text. The measurements were performed at 100 K.

**Supplementary reference**

[S1] K. M. Fijalkowski, N. Liu, M. Hartl, M. Winnerlein, P. Mandal, A. Coschizza, A. Fothergill, S. Grauer, S. Schreyeck, K. Brunner, M. Greiter, R. Thomale, C. Gould, and L.W. Molenkamp, *Phys. Rev. B Condens. Matter* **2021,** 103, 235111